# Time evolution of the fission-decay width under the influence of dissipation[§]

B. Jurado[a,#], K.-H. Schmidt[a], J. Benlliure[b]

[a] *Gesellschaft für Schwerionenforschung, Planckstraße 1, D-64291 Darmstadt, Germany*
[b] *Faculdad de Fisica, Universidad de Santiago de Compostela, E-15706 Santiago de Compostela, Spain*

**Abstract**
Different analytical approximations to the time-dependent fission-decay width used to extract the influence of dissipation on the fission process are critically examined. Calculations with a new, highly realistic analytical approximation to the exact solution of the Fokker-Planck equation sheds doubts on previous conclusions on the dissipation strength made on the basis of less realistic approximations.

Viscosity is a fundamental collective property of nuclei that is scarcely known, in spite of intense experimental and theoretical efforts. The absolute magnitude as well as the dependences on temperature and deformation are still subject of debate [1, 2]. The observation of dissipative effects in fission provides a very powerful tool to extract the relevant experimental information. However, the model calculations applied for interpreting these results are often not satisfactory. The present publication provides an improved treatment of the time evolution of the fission-decay width under the influence of dissipation and demonstrates the application to a set of experimental data.

The modeling of the fission-decay width at high excitation energies requires the treatment of the evolution of the fission degree of freedom as a dissipative process, determined by the interaction of the fission collective degree of freedom with the heat bath formed by the individual nucleons [3, 4]. Such a process can be described by the Fokker-Planck equation (FPE) [5], where the variable is the time- and dissipation-dependent probability distribution $W(x, p; t, \beta)$ as a function of the deformation in fission direction $x$ and its canonically conjugate momentum $p$. $\beta$ is the reduced dissipation coefficient. The solution of the FPE leads to a time-dependent fission-decay width $\Gamma_f(t)$, whose shape is influenced by the dissipation strength.

Grangé, Jun-Qing and Weidenmüller [4] solved numerically the time-dependent FPE for a nucleus with a certain intrinsic excitation energy and with the phase space initially populated according to the zero-point motion. In order to systematically study the influence of different parameters such as fission barrier, nuclear temperatures and dissipation strength on the fission-decay width $\Gamma_f(t)$, we have repeated such calculations[*] using a deformation-dependent nuclear potential given by two smoothly joined parabolas with opposite curvatures. The absolute values of the curvatures in the ground state and at the barrier have been taken from ref. [6]. The obtained time-dependent fission-decay width in the case of $^{238}$U at a temperature of 3 MeV for different values of the reduced dissipation coefficient $\beta$ is shown in figure 1 with solid lines.

---

[§] This work forms part of the PhD thesis of B. Jurado
[#] Corresponding author: B. Jurado, GANIL, Boulevard Henri Becquerel, B.P.5027, 14076 Caen CEDEX 5, France, e-mail: jurado@ganil.fr (present address)
[*] The numerical solutions of the FPE were obtained with the software package FEMLAB (Comsol AB, Stockholm).

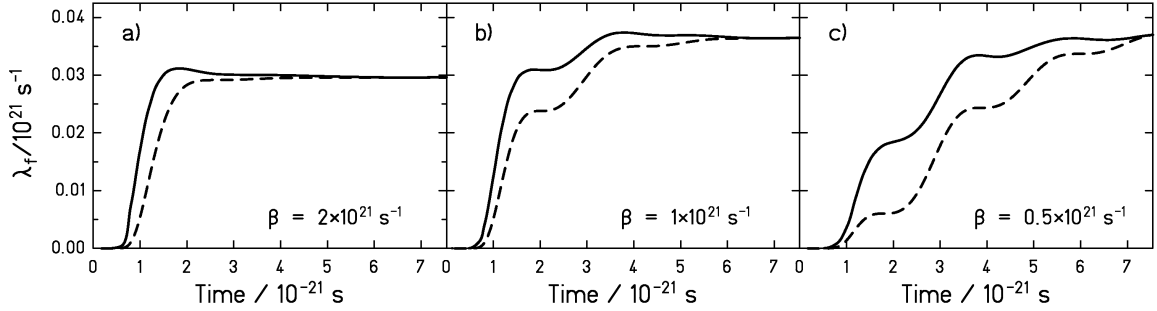

Figure 1: Fission rate $\lambda_f(t) = \Gamma_f(t)/\hbar$ as a function of time for three different values of the reduced dissipation coefficient $\beta$ for $^{238}$U at a temperature $T = 3$ MeV. The solid line is the numerical solution of the FPE while the dashed line is calculated using equation (8) with $W_n(x=x_b;t,\beta)$ taken from equations (9) and (10). The initial conditions correspond to the zero-point motion (see text).

However, these numerical calculations are too time consuming to be implemented in nuclear-reaction codes, which have to be used in order to extract the relevant physical information from available experimental data. The same problem holds for the application of the Langevin equation, which has been developed as an alternative possibility [7, 8]. Therefore, in the model calculations usually one of the following approximations [9, 10] for the time-dependent fission-decay width $\Gamma_f(t)$ is applied:

- A step function [9] that sets in at time $\tau_f$: $\Gamma_f(t) = \begin{cases} 0, t < \tau_f \\ \Gamma_f^k, t \geq \tau_f \end{cases}$ (1)

- An exponential in-growth function [10]: $\Gamma_f(t) = \Gamma_f^k \{1 - \exp(-t/\tau)\}$ (2)

where $\tau = \tau_f / 2.3$, $\tau_f$ is the transient time, defined [11] as the time in which $\Gamma_f(t)$ reaches 90 % of its asymptotic value given by the Kramers fission-decay width $\Gamma_f^k$ [3].

$$\Gamma_f^k = \Gamma_f^{BW} \cdot \left[\left(1+\left(\frac{\beta}{2\omega_0}\right)^2\right)^{1/2} - \frac{\beta}{2\omega_0}\right]$$ (3)

where $\Gamma_f^{BW}$ is the fission-decay width given by the transition-state model [12], and $\omega_0$ is the frequency of the oscillator, corresponding to the inverted potential at the barrier.

These approximations strongly deviate from the numerical solution and thus may severely influence the results [13, 14]. As an example, table 1 summarizes the results of different calculations in comparison with the experimental fission cross section after nuclear excitations of the system $^{238}$U + Pb at 1 $A$ GeV from ref. [17]. Obviously, the reduced dissipation coefficient that describes the experimental result varies by a factor of two between $2 \cdot 10^{21}$ s$^{-1}$ and $4 \cdot 10^{21}$ s$^{-1}$, depending on the approximation used. In order to come to a reliable conclusion, the calculation should be repeated using a more realistic description of the time-dependent fission-decay width. This is the aim of the present work. The influence of other model parameters used in the calculation and a discussion on the validity of the initial conditions will be presented in another publication [14].

An analytical approximate solution to the FPE for a realistic nuclear potential was obtained in the case of the over-damped motion ($\beta > 2 \cdot 10^{21}$ s$^{-1}$) [11]. However, this solution did not fulfil the requirement of the exact solution of the FPE to vanish at the beginning of the deexcitation process [4].

We have developed a new highly realistic description of the fission-decay width [2] based on the analytical solution of the FPE when the nuclear potential is approximated by a parabola.
The time-dependent fission-decay width is defined as [4]:

$$\Gamma_f(t) = \hbar \lambda_f(t) = \hbar \frac{\int_{-\infty}^{+\infty} v W(x = x_b, v; t, \beta) dv}{\int_{-\infty}^{x_b} \int_{-\infty}^{+\infty} W(x, v; t, \beta) dv dx} \quad (4)$$

where $\lambda_f(t)$ is the fission rate, $x_b$ is the deformation at the barrier, $v$ is the velocity ($v=dx/dt$) and $W(x,v;t,\beta)$ is the probability distribution. The denominator measures the part of the probability distribution still caught inside the fission barrier. Due to the flux over the fission barrier, its value gradually decreases.

Table 1: Experimental total nuclear-induced fission cross section of $^{238}$U(1 $A$ GeV) in a lead target, compared with different calculations performed with the code ABRABLA [15, 16]. The experimental cross sections are taken from [17]. The simultaneous break-up stage as described in ref. [18] is included in the calculation, so that a limit for the initial temperature of 5.5 MeV of the sequential decay (evaporation cascade) is imposed. A first set of calculations was performed with the transition-state model [12]. The other calculations were performed with different descriptions of $\Gamma_f(t)$ and different values of $\beta$. The last calculation was performed with $\Gamma_f(t)$ as the analytical solution of the FPE given by equation (8) with $W_n(x=x_b;t,\beta)$ taken from equations (9) and (10).

| Experiment | | (2.07 ± 0.38) b |
|---|---|---|
| Transition-state model | | 3.33 b |
| $\Gamma_f(t)$ step (ref. [9]) | $\beta = 2 \cdot 10^{21}$ s$^{-1}$ | 2.00 b |
| $\Gamma_f(t)$ step (ref. [9]) | $\beta = 4 \cdot 10^{21}$ s$^{-1}$ | 1.54 b |
| $\Gamma_f(t) \sim 1 - \exp(-t/\tau)$ (ref. [10]) | $\beta = 2 \cdot 10^{21}$ s$^{-1}$ | 2.52 b |
| $\Gamma_f(t) \sim 1 - \exp(-t/\tau)$ (ref. [10]) | $\beta = 4 \cdot 10^{21}$ s$^{-1}$ | 2.04 b |
| $\Gamma_f(t)$ FPE (this work) | $\beta = 2 \cdot 10^{21}$ s$^{-1}$ | 2.09 b |

We define the normalized probability distribution at the barrier deformation $x_b$, $W_n(x = x_b, v; t, \beta)$, as:

$$W_n(x = x_b, v; t, \beta) = \frac{W(x = x_b, v; t, \beta)}{\int_{-\infty}^{x_b} \int_{-\infty}^{+\infty} W(x, v; t, \beta) dv dx} \quad (5)$$

Considering equation (5) and taking into account the stationary value $\Gamma_f^K$, the fission-decay width of equation (4) can be reformulated as:

$$\Gamma_f(t) = \frac{\int_{-\infty}^{+\infty} v W_n(x=x_b, v; t, \beta) dv}{\int_{-\infty}^{+\infty} v W_n(x=x_b, v; t \to \infty, \beta) dv} \Gamma_f^K \qquad (6)$$

At this point we introduce two approximations. First we consider that the shape of the probability distribution at the barrier deformation $W(x = x_b, v; t, \beta)$ as a function of the velocity $v$ is constant in time, and only its height varies with time as $C(t)$:

$$W(x=x_b, v; t, \beta) \approx C(t) \cdot W(x=x_b, v; t \to \infty, \beta) \qquad (7)$$

This statement is valid in the over-damped motion were the equilibrium in velocity is established very rapidly. However, this is still applicable somewhat outside the over-damped regime, since $x_b$ is far away from the initial deformation. Thus, the time needed for the probability distribution to reach the fission barrier is long enough for the velocity to equilibrate.

As a consequence, we can express the fission-decay width in the following form, which has no explicit dependence on velocity:

$$\Gamma_f(t) \approx \frac{W_n(x=x_b; t, \beta)}{W_n(x=x_b; t \to \infty, \beta)} \Gamma_f^K \qquad (8)$$

The second approximation consists in using for $W_n(x = x_b; t, \beta)$ the solution of the FPE obtained using a parabolic nuclear potential [19] for zero deformation and zero velocity as initial conditions:

$$W_n(x=x_b; t, \beta) = W^{par}(x=x_b; t, \beta) = \frac{1}{\sqrt{2\pi} \cdot \sigma} \exp\left(-\frac{x_b^2}{2\sigma^2}\right) \qquad (9)$$

where $\sigma^2$ is a time-dependent function of the form:

$$\sigma^2 = \frac{T}{\mu\omega_1^2}\left\{1 - \exp(-\beta\, t) \cdot \left[\frac{2\beta^2}{\beta_1^2}\sinh^2\left(\frac{\beta_1 t}{2}\right) + \frac{\beta}{\beta_1}\sinh(\beta_1 t) + 1\right]\right\} \qquad (10)$$

where $T$ is the nuclear temperature, $\mu$ is the reduced mass associated to the deformation degree of freedom, $\omega_1$ describes the curvature of the potential at the ground state and $\beta_1 = (\beta^2 - 4\omega_1^2)^{1/2}$.

Implementing equations (9) and (10) in equation (8) results in an analytical expression for $\Gamma_f(t)$. As initial condition of the problem, we have chosen the zero-point motion at the ground-state deformation, which is adequate [13] to the reaction listed in table 1. The zero-point motion was taken into account by shifting the time scale by a certain amount: For the under-damped case ($\beta < 2 \cdot 10^{21}$ s$^{-1}$), the deformation and the momentum coordinate saturate at about the same time. Therefore, the time shift needed for the probability distribution to reach the width of the zero-point motion in deformation space is equal to the time that the average energy of the collective degree of freedom needs to reach the value $\frac{1}{2}\hbar\omega_1$ associated to the zero-point motion:

$$t_0 = \frac{1}{\beta} \ln\left(\frac{2T}{2T - \hbar\omega_1}\right) \tag{11}$$

In the over-damped regime ($\beta \geq 2\cdot 10^{21}$ s$^{-1}$), the momentum coordinate saturates very fast, while the population of the deformation space is a diffusion process. In this case, the time in which the variance in the deformation coordinate of the probability distribution acquires the corresponding value of the zero-point motion follows from the solution of the Fokker-Planck equation [16]:

$$t_0 = \frac{\hbar\beta}{4\omega_1 T} \tag{12}$$

if the influence of the potential on the diffusion process is neglected, which is anyhow small in the range of the zero-point motion. In this way, the zero-point motion is taken as the initial condition in our calculations throughout this paper, in particular in the approximate description of the time-dependent fission-decay width of figure 1. For the numerical calculations, however, the exact distributions in position and velocity of the zero-point motion were taken.

In addition, the reduced mass was obtained using the relation:

$$\mu\omega_1^2 = 2K \tag{13}$$

where $K$ is the stiffness of the potential [6] and $\hbar\omega_1 = 1$ MeV [9]. The deformation at the barrier was obtained using the expression of reference [20]

$$x_b = \frac{7}{3}y - \frac{938}{765}y^2 + 9.499768 y^3 - 8.050944 y^4 \tag{14}$$

where $y = 1-\alpha$ and $\alpha$ is the fissility parameter.

As can be seen in figure 1a), this analytical approximation quite well reproduces the exact solution for the critical damping ($\beta = 2\cdot 10^{21}$ s$^{-1}$). A similar agreement is obtained in the over-damped regime ($\beta > 2\cdot 10^{21}$ s$^{-1}$) [14]. The approximation also gives a rather good description of the slightly under-damped motion ($\beta = 1\cdot 10^{21}$ s$^{-1}$), shown in figure 1b). Even in the strongly under-damped case ($\beta = 0.5\cdot 10^{21}$ s$^{-1}$), the onset of the fission decay and the oscillations are reproduced very well as demonstrated in figure 1c), although the absolute magnitude of the fission rate is somewhat underestimated.

Using the new description of the time-dependent fission-decay width, given by equations (8), (9) and (10), another model calculation has been performed. The result is listed in the last line of table 1. It is interesting to note that the result is rather close to the calculation performed with the step function, while it strongly deviates from the calculation with the exponential-type in-growth function. Obviously, the total suppression of the fission-decay width at the beginning of the deexcitation cascade is an essential feature of a realistic description of the influence of dissipation on the fission process. This makes also clear that the quantitative deviations of the new approximate description from the numerical solution of the time-dependent fission-decay width are not crucial and will not lead to substantially different results.

The total suppression of the fission-decay width for small time values and the gradual increase are the most critical features of a realistic formulation of the time-dependent fission-decay width [2, 13] in reactions that pass by a compound nucleus with small shape distortions. These features, which were quantitatively or even qualitatively missed in the previously used descriptions, are well reproduced by the analytical approximation proposed in the present work. Using the new description of the time-dependent fission-decay width in nuclear-model codes will improve the quality of such calculations appreciably and allow for more realistic conclusions on the dissipation strength when interpreting experimental data. In particular, any conclusions previously drawn from calculations using the exponential-like in-growth function should be re-examined.

## Acknowledgement

We thank O. Yordanov for the assistance in using the FEMLAB code as well as A. Kelić and A. Heinz for reading the manuscript. This work has partly been supported by the European Union under contract number EC-HPRI-CT-1999-00001 and under the ECT* 'STATE' contract.

## References:

[1] D. Hilscher, H. Rossner, Ann. Phys. Fr. **17** (1992) 471.
[2] B. Jurado, 2002, PhD thesis, University Santiago de Compostela.
[3] H. A. Kramers, Physika **VII 4** (1040) 284.
[4] P. Grangé, L. Jun-Qing, and H. A. Weidenmüller, Phys. Rev. **C 27** (1983) 2063.
[5] The Fokker-Planck Equation, H. Risiken, Springer, Berlin (1989), ISBN 0-387-50498-2.
[6] W. D. Myers and W. Swiatecki, Nucl. Phys. **A 81** (1966) 1.
[7] Y. Abe et al. Phys. Rep. **275** (1996) 49.
[8] P. Fröbrich and I. I. Gontchar, Phys. Rep. **292** (1998) 131.
[9] E. M. Rastopchin et al., Yad. Fiz. **53** (1991) 1200 [Sov. J. Nucl. Phys. **53** (1991) 741].
[10] R. Butsch et al., Phys. Rev. **C 44** (1991) 1515.
[11] K.-H. Bhatt, P. Grangé, and B. Hiller, Phys. Rev. **C 33** (1986) 954.
[12] N. Bohr and J. A. Wheeler, Phys. Rev. **56** (1939) 426.
[13] B. Jurado et al., XXXIX International Winter Meeting on Nuclear Physics, January 22-26 2001, Bormio, Italy.
[14] B. Jurado, K.-H. Schmidt, J. Benlliure, and A. R. Junghans, submitted to Nucl. Phys. **A**.
[15] J.-J. Gaimard and K.-H. Schmidt, Nucl. Phys. **A 531** (1991) 709.
[16] A. R. Junghans et al., Nucl. Phys. **A 629** (1998) 635.
[17] Th. Rubehn et al., Phys. Rev. **C 53** (1996) 3143.
[18] K.-H. Schmidt, M. V. Ricciardi, A. Botvina, and T. Enqvist, Nucl. Phys. **A 710** (2002) 157.
[19] S. Chandrasekhar, Rev. Mod. Phys. **15** (1943) 1.
[20] R. W. Hasse and W. D. Myers, "Geometrical relationships of Macroscopic Nuclear Physics" Springer-Verlag Berlin Heidelberg (1988) ISBN 3-540-17510-5.